\documentclass[12pt]{article}

\usepackage{hyperref,amsfonts,amssymb, amsmath,theorem,mathrsfs}
\usepackage[all]{xy}

\usepackage[english]{babel}

\DeclareMathAlphabet{\bfit}{OT1}{cmr}{bx}{it}
\DeclareMathAlphabet{\mathpzc}{OT1}{pzc}{m}{it}

\def\A{\mathcal A}
\def\B{\mathcal B}
\def\H{\mathcal H}
\def\Q{\mathbf Q}
\def\K{\mathcal K}

\def\l{\lambda}

\def\vk{\varkappa}

\def\dsize{\displaystyle}
\def\fs{\footnotesize}

\def\nn{\nonumber }
\def\p{\partial }

\def\bq{ \begin{equation} }
\def\eq{ \end{equation} }
\def\ben{ \begin{eqnarray} }
\def\en{ \end{eqnarray} }
\def\ba{ \begin{array} }
\def\ea{ \end{array} }

\def\on#1#2{\mathop{\vbox{\ialign{##\crcr\noalign{\kern2pt}
$\scriptstyle{#2}$\crcr\noalign{\kern2pt\nointerlineskip}
\kern-2pt$\hfil\displaystyle{#1}\hfil$\crcr}}}\limits}

\newtheorem{prop}{Proposition}

\theorembodyfont{\rm}
\newtheorem{rem}{Remark}

\begin{document}

%%%%%%%%%%%%%%%%%%%%%%%%%%%%%%%%%%%%%%

\baselineskip=15pt

\vspace{1cm} \noindent {\LARGE \textbf{On integration of the
Kowalevski gyrostat and the Clebsch problems}} \vskip1cm \hfill
\begin{minipage}{13.5cm}
\baselineskip=15pt {\bf I V Komarov
 and
 A V Tsiganov }\\ [2ex]
{ V.A. Fock Institute of Physics,
St.Petersburg State University,\\
St.Petersburg, Russia\\
}

 \vskip1cm{\bf Abstract}
%\baselineskip=3D15pt

For the Kowalevski gyrostat change of variables similar to that of the
Kowalevski top is done. We establish one to one correspondence
between the Kowalevski gyrostat and the Clebsch system and
demonstrate  that Kowalevski variables for the gyrostat
practically coincide with elliptic coordinates on sphere for the
Clebsch case. Equivalence of considered integrable systems allows
to construct two Lax matrices for the gyrostat  using known
rational and elliptic Lax matrices for the Clebsch model.
Associated with these matrices solutions of the Clebsch system
and, therefore, of the Kowalevski gyrostat problem are discussed.
The K\"otter solution of the Clebsch system in modern notation is
presented in detail.

\end{minipage}
\vskip0.8cm \noindent{ PACS numbers: 02.30.Ik, 02.30.Uu, 02.30.Zz,
02.40.Yy, 45.30.+s } \vglue1cm \textbf{Corresponding Author}: A V
Tsiganov, St.Petersburg State University, St.Petersburg, Russia,
E-mail: tsiganov@mph.phys.spbu.ru \newpage

\section{Introduction}

Kowalevski gyrostat is a one parameter integrable extension of the
Kowalevski top constructed in 1987 \cite{gyro-k}, \cite{gyro-y}. For
the Kowalevski gyrostat the Lax representation with spectral
parameter was found in framework of a general group-theoretical
approach to integrable systems using the Lie algebras $so(3,2)$ or
$sp(4)$ \cite{RS,brs89}. For the Kowalevski top the spectral curve
generated by this Lax matrix differs from the original Kowalevski
curve.

Solutions for the top in terms of the Prym theta-functions were
obtained by finite-band integration technique in \cite{brs89}, where
it was said that the gyrostat can be integrated in a similar way. We
do not know separated variables and separated equations associated
with the Lax matrices \cite{brs89} neither for the top nor for the
gyrostat.

Another algebro-geometric approach to study of the Kowalevski top
was developed in \cite{HH,avm88}. In this method detailed analysis
of the level surfaces of constant of motion allows to establish a
birational isomorphism between the Kowalevski flow and the flows of
other integrable systems that are linearizable on abelian varieties
of the same type. In \cite{HH} the Kowalevski top is related with
the Neumann system and in \cite{avm88} with the Schottky-Manakov top
on $so(4)$. The known Lax matrices and separated variables for these
systems give rise to the Lax matrices and the separated variables
for the Kowalevski top.

The aim of this paper is to extend the Kowalevski treatment of the
top to the gyrostat and to construct an isomorphism between the
Kowalevski gyrostat and the Clebsch case of the motion of a rigid
body in ideal fluid. As a byproduct one gets new rational and
elliptic Lax matrices for the Kowalevski gyrostat together with the
corresponding integration procedures.

\section{The Kowalevski gyrostat}
\setcounter{equation}{0}

 Let two vectors $\bfit J$ and $\bfit
x$ are coordinates on the phase space $M$. As a Poisson manifold $M$
is identified with Euclidean algebra $e(3)^*$ with the Lie-Poisson
brackets
\begin{equation}\label{e3}
\,\qquad \bigl\{J_i\,,J_j\,\bigr\}=\varepsilon_{ijk}J_k\,, \qquad
\bigl\{J_i\,,x_j\,\bigr\}=\varepsilon_{ijk}x_k \,, \qquad
\bigl\{x_i\,,x_j\,\bigr\}=0\,,
\end{equation}
where $\varepsilon_{ijk}$ is the totally skew-symmetric tensor.
These brackets have two Casimir functions
\begin{equation}\label{caz0}
A=\bfit x^2\equiv\sum_{k=1}^3 x_k^2, \qquad
  B=(\bfit x \cdot \bfit
J)\equiv\sum_{k=1}^3 x_kJ_k .
\end{equation}
Fixing their values one gets a generic symplectic leaf of $e(3)$
$$
{\mathcal O}_{ab}:\qquad \{{\bfit x}\,, {\bfit J}\,:~A=a,~~
B={b}\}\,,
$$
which is a four-dimensional symplectic manifold.

The Hamilton function for the original Kowalevski top is given by
\begin{equation}\label{H-Kow}
H_{\mbox{\fs top}}=\frac12(J_1^2+J_2^2+ 2J_3^2)+c x_1\,,\qquad
c\in\mathbb C.
\end{equation}
This Hamiltonian and additional integral of motion
\bq\label{K-Kow} K_{\mbox{\fs top}}=\xi_{1}\,\cdot \xi_{2}, \eq
are in  involution and define a moment map whose fibers are
Liouville tori in ${\mathcal E}_{ab}$. Here \bq\label{xi12}
  \xi_{1}=z_{1}^2-2c(x_1+ix_2)\,, \qquad \xi_{2}=z_{2} ^2-2c (x_1-ix_2)
\eq
and
\bq\label{z12}
z_1=J_1 + i J_2, \qquad z_2=J_1-i J_2.
\eq

The Kowalevski gyrostat \cite{gyro-k}, \cite{gyro-y} is an
integrable extension of the corresponding top
 defined by the following constants of motion
\ben
 \label{HGSK}
H&=&H_{\mbox{\fs top}}-\lambda J_3=\frac12(J_1^2+J_2^2+
2J_3^2-2\lambda J_3 )+c x_1\,,\\
\nn\\
\label{KGSK} K&=&\xi_{1} \xi_{2}+4\lambda\Bigl((J_3-\lambda)z_{1}
z_{2} -(z_{1} +z_{2} )c x_3\Bigr)
\en
in involution $\{H\,,K\}=0$. The gyrostat generalization of the
Kowalevski top is essential because the corresponding additional
terms in the Hamiltonian mimic quantum corrections to the top
\cite{gyro-k}.

The equations of motion are given by the customary Euler-Poisson
equations
\bq\label{Kow-eqm}
\
 X:\qquad \dot{\bfit J}=\bfit J\times\frac{\partial H}{\partial \bfit
J}
 +\bfit x\times\frac{\partial H}{\partial \bfit x},\qquad
\dot{\bfit x}=\bfit x\times \frac{\partial H}{\partial \bfit J}\,,
\eq
where $\bfit x\times \bfit z$ means cross product of two vectors.
Equations (\ref{Kow-eqm}) can be rewritten as
\[
X\left(
\begin{array}{c}
 \bfit J \\
 \bfit x \\
\end{array}
\right)=P_0\,dH\,,\qquad P_0=\left(%
\begin{array}{cc}
 \mathbf J & \mathbf X \\
 -\mathbf X & 0
\end{array}%
\right),
\]
where
\[
\mathbf J=\left(\begin{array}{ccc}
           0 & J_3 & -J_2 \\
           -J_3 & 0 & J_1 \\
           J_2 & -J_1 & 0
          \end{array}\right)\,,\qquad
\mathbf X=\left(\begin{array}{ccc}
           0& x_3 & -x_2 \\
            -x_3 & 0 & x_1 \\
           x_2 & -x_1 & 0
          \end{array}\right)\,.\]
We do not show the second commuting flow for brevity.

%The bi-hamiltonian structure of the Kowalevski top was discussed in \cite{mar98}.

\begin{rem} \cite{gyro-k}: Additional terms in $K$ are closely related with the third order in momenta integral of the Goryachev-Chaplygin gyrostat
\[
G=2((J_3-2\lambda)z_{1} z_{2} +(z_{1} +z_{1} )c x_3)
\]
and expressed in terms of the Poisson brackets of $\xi_1, \xi_2$
\bq
K=K_{\mbox{\fs top}}+i\l \{\xi_1, \xi_2\}-16\l^2 z_{1} z_{2} \, .
\eq
\end{rem}

\section{The Kowalevski variables}
\setcounter{equation}{0}

We will introduce variables $s_{1,2}$ for the Kowalevski gyrostat
step by step following original papers \cite{kow89} and
\cite{kotter} where the separation of variables for the top was
constructed.

At first we made a transition from initial variables to new
variables $\xi_{1,2}$ (\ref{xi12}), $z_{1,2}$ (\ref{z12}) and
organize four constants of motion  in the following matrix identity
\[
\left(\begin{array}{ll}
 4H & 4cB\\
 4cB & 4c^2A-K\\
\end{array}%
\right)=4\left(\begin{array}{cc}
J_3^2 & c x_3 J_3 \\[2mm]
c x_3 J_3 & c^2 x_3^2
\end{array}\right)
-4\l \left(\begin{array}{cc}
J_3 & 0 \\[2mm]
0 & z_{1}z_{2}(J_3-\l)-c(z_{1}+z_{2})x_3
\end{array}\right)+
\]
\bq
+ \left(\begin{array}{cc}
(z_1+z_2)^2 & z_1z_2 (z_1+z_2) \\[2mm]
 z_1z_2(z_1+z_2) & z_1^2 z_2^2
\end{array}\right)
- \left(\begin{array}{cc}
\xi_1+\xi_2 & \xi_1 z_2+\xi_2 z_1 \\[2mm]
\xi_1 z_2+\xi_2 z_1 & \xi_1 z_2^2+\xi_2 z_1^2
\end{array}\right)
\label{D-identity} \eq The second step consists of exclusion two
variables $x_3$ and $J_3$ using velocities $\dot z_{i}=\{H, z_{i}\}$
\bq \label{x3-J3}
x_3=\frac{i}{c}\frac{\dot{z}_1z_2+z_1\dot{z}_2}{z_1-z_2}\,,\qquad
J_3=\frac{i(\dot{z}_2+\dot{z}_1)}{(z_1-z_2)}+\l\,.
\eq
Similarity transform $U^{t} (~\cdot~) U$ of the both sides of
(\ref{D-identity}) with auxiliary matrix $U$
\bq
U=\left(\begin{array}{cc}
z_1 & z_2 \\[2mm]
-1 & -1
\end{array}\right), \qquad
U^{t}=\left(\begin{array}{cc}
z_1 & -1 \\[2mm]
z_2 & -1
\end{array}\right)
\eq
brings us to the following matrix identity for the gyrostat
\ben
4\left(\begin{array}{cc}
 \dot z_{1}^2 & - \dot z_{1} \dot z_{2}\\[2mm]
 - \dot z_{1} \dot z_{2} & \dot z_{2}^2
\end{array}\right)
&+&4 i \l(z_{1} -z_{2} ) \left(\begin{array}{cc}
 \dot z_{1} & 0 \\[2mm]
 0 &   \dot z_{2}
\end{array}\right)
+(z_{1} -z_{2} )^2 \left(\begin{array}{cc}
\xi_{1}& -2H \\[2mm]
-2H & \xi_{2}
\end{array}\right)
\nn\\
\label{mat-id}\\
&-&\left(\begin{array}{cc}
R(z_1, z_1) & R(z_1, z_2) \\[2mm]
R(z_1, z_2) & R(z_2, z_2)
\end{array}\right) = 0
.\nn
\en
Here
\bq
\label{R12} R(z_{1} ,z_{2} )=
 z_{1} ^2 z_{2} ^2-2H(z_{1} ^2+z_{2} ^2) -4c B(z_{1} +z_{2} )-4 c^2\,
A+K.
\eq
The diagonal entries of the identity (\ref{mat-id}) allows to
express variables $\xi_{1,2}$ as
\[
\xi_{k}=\frac{4i\lambda\dot{z}_{k}}{z_1-z_2}-\frac{4\dot{z}_{k}^2-R(z_k,z_k)}{(z_1-z_2)^2},\qquad
k=1,2,
\]
and one gets integrals $H$ and $K$ in terms of biquadratic
polynomial $R$ (\ref{R12}) and two pairs of Lagrangian variables
$z_{1,2}$ and $\dot{z}_{1,2}$
\ben H&=&
-\dfrac{4\dot{z}_1\dot{z}_2+R(z_1,z_2)}{2(z_1-z_2)^2}\,,
\label{HKz}\\
\nn\\[2mm]
K&=&- \frac{16\dot{z}_1\dot{z}_2}{(z_1-z_2)^2}\lambda^2
-4i\lambda\Bigl(\dot{z}_1\dfrac{\p}{\p z_1}-\dot{z}_2\dfrac{\p}{\p
z_2}\Bigr)\dfrac{R(z_1,z_2)}{(z_1-z_2)^2}\nn\\[2mm]
&&+\frac{(4\dot{z}_1^2-R(z_1,z_1))(4\dot{z}_2^2-R(z_2,z_2))}{(z_1-z_2)^4}\,.
\label{KKz}
\en
On the level surface of integrals of motion
\bq\label{Int-Surf}
\Sigma=\left\{A=a,~ B=b,~ H=h,~ K=k \right\}
\eq
relations (\ref{HKz}) and (\ref{KKz})
\[\left.\Phi_{1,2}(z_1,z_2,\dot{z}_1,\dot{z}_2,A,B,H,K)\right|_\Sigma=0\]
can be considered as equations of motion determining  two
dimensional dynamical system. Unfortunately variables $z_{1,2}$ do
not commute $\{z_1, z_2 \}\ne 0$, so one has to look for more
convenient parametrization.

\begin{rem}\label{Weil-rem} Associated with the
 fourth degree polynomials $R(z_k,z_k)$ (\ref{R12})
\[
R(z_k,z_k)=a_0z_k^4+4a_1z_k^3+6a_2z_k^2+4a_3z_k+a_4\,,\qquad
a_i\in\mathbb R
\]
differential equations
\bq\label{Eul-mEq}
 \dfrac{\dot{z}_1}{\sqrt{R(z_1,z_1)}}=\pm\dfrac{\dot{z}_2}{\sqrt{R(z_2,z_2)}}\,,
\eq originally appeared in the Euler studies of equation of
lemniscate and invariance of the corresponding elliptic integrals
\cite{eul68}. In particular Euler proved that equations
(\ref{Eul-mEq}) have an algebraic integral
\bq
\label{Q-Kow}
\mathcal E(z_1,z_2,s)=(z_1-z_2)^2\,s^2-R(z_1,z_2)\,s+W=0\,, \eq
where $R(z_1,z_2)$ is a mixed biquadratic form similar to
(\ref{R12})
\bq
\label{quEuler}
R(z_1,z_2)=a_0z_1^2z_2^2+2a_1z_1z_2(z_1+z_2)+3a_2(z_1^2+z_2^2)+2a_3(z_1+z_2)+a_4,
\eq and
\[
W=\dfrac{R(z_1,z_2)^2-R(z_1,z_1)R(z_2,z_2)}{4(z_1-z_2)^2}.
\]

In algebro-geometric terms \cite{we83}, Euler studied automorphisms
$(u_1,z_1)\to(u_2,z_2)$ of the algebraic curve of genus one
\bq
\label{curv4} \mathcal C:\quad u^2=R(z,z),
\eq
which change a sign of the corresponding holomorphic form
$dz/u\to\pm dz/u$. Thus every algebraic curve of genus one is
isomorphic to a complex torus (cubic elliptic curve), which
equivalent to Jacobian of $\mathcal C$. These automorphisms are
parameterized by points of a smooth elliptic curve
\bq
\label{curv3} \Gamma:\quad \eta^2=P_3(s),\qquad
P_3(s)=4s^3+g_1s^2+g_2s+g_3\,,
\eq
where $g_k$ are functions on initial parameters $a_0,\ldots,a_4$.
According to Weil \cite{we83}, if $O_k=(u_k,z_k)$, $k=1,2$, are two
points of $\mathcal C$ and $N_k=(\eta_k,s_k)$, $k=1,2$, denote two
points of $\Gamma$ related by $O_1=N_1+N_2$ and $O_2=N_1-N_2$ then
\bq \label{Eul-We}
\dfrac{dz_1}{u_1}+\dfrac{dz_2}{u_2}=\dfrac{ds_{1}}{\eta_{1}},\qquad
\dfrac{dz_1}{u_1}-\dfrac{dz_2}{u_2}=\dfrac{ds_{2}}{\eta_{2}}.
\eq
It is infinitesimal version of the Weil interpretation of the Euler
results. These results are independent on the choice of affine
coordinates $(u,z)$ and $(\eta,s)$ on the curves $\mathcal C$ and
$\Gamma$, respectively.
\end{rem}

The third step of Kowalevski in \cite{kow89} is to apply
automorphisms of auxiliary elliptic curve (\ref{curv4}) given  in
Remark 2 by introduction her famous variables $s_{1,2}$
\bq
\label{s1,2} s_{1,2}=\frac{R(z_{1} ,z_{2} ) \pm\sqrt{R(z_{1} ,z_{1}
)R(z_{2} ,z_{2} )}} {2(z_{1} -z_{2} )^2},
\eq
which are transcendental integrals of the corresponding Euler
equations (\ref{Eul-mEq}) \cite{eul68}.

The variables $s_{1,2}$ are eigenvalues of an auxiliary spectral
problem \bq \label{aux-evp} \left(\begin{array}{cc}
R(z_1, z_1) & R(z_1, z_2) \\[2mm]
R(z_1, z_2) & R(z_2, z_2)
\end{array}\right){\Psi}=2\, s~ \sigma_1 (z_1 - z_2)^2
 \Psi , \qquad
\sigma_1= \left(\begin{array}{cc}
0& 1 \\[2mm]
1 & 0
\end{array}\right),
\eq
that is naturally extracted from (\ref{mat-id}). Its characteristic
polynomial
\bq
\label{Es1,2} {\mathcal E}(s)=(z_1-z_2)^2\,(s-s_1)(s-s_2)\,.
\eq
coincides with the Euler algebraic integral (\ref{Q-Kow}). For its
analysis see, e.g., Golubev \cite{Golubev}.

The matrix of eigenfunctions $\Psi$ of the spectral problem
(\ref{aux-evp}) reads
\bq
\label{Psi} \Psi=\left(\begin{array}{cc}
\dsize \frac {1}{\sqrt{R(z_1, z_1)}} &\dsize \frac {1}{\sqrt{R(z_2, z_2)}} \\[2mm]
\dsize -\frac {1}{\sqrt{R(z_1, z_1)}} &\dsize \frac {1}{\sqrt{R(z_2,
z_2)}}
\end{array}\right)\,.
\eq

The idea of Kowalevski to pass to new variables $s_{1,2}$ is
appeared to be very fruitful in her treatment of the top. For the
Kowalevski gyrostat as well as for the top these variables have the
following main property:

\begin{prop} Functions
$s_{1,2}$ (\ref{s1,2}) are Poisson commute $\{s_1,\, s_2\}=0.$
\end{prop}
For $\lambda=0$ the straightforward proof may be founded in
\cite{komkuz1,ves2}. For $\lambda\neq 0$ this unexpected and crucial
observation was obtained by direct calculation of the Poisson
brackets. It allows to suggest that for the gyrostat variables
$s_{1,2}$ provide an essential step to separation of variables  and
this stimulated us to write down evolutionary equations and the
integrals of motion for it in terms of $s_{1,2}$ and their
velocities $\dot s_{1,2}$ .

\begin{rem}\label{Kow-rem} In her original paper Kowalevski
uses also another preferred coordinate system $(t,w)$ in which curve
$\Gamma$ (\ref{curv3}) has a Weierstrass normal form associated with
equation $P_3(w)= 4w^3+g_2w+g_3$. i.e. with $g_1=0$ (see
\cite{kow89}, p. 188). These variables $w_i=s_i-2H$ do not Poisson
commute. The counterpart of (\ref{mat-id}) in $w_{1,2}$ variables
gives as its off diagonal entries a nonphysical identity for $w_1,
w_2$ and $\dot w_1, \dot w_2$ instead of (\ref{Hs}).
\end{rem}

Function ${\cal E}(z_1, z_2, s)$ (\ref{Q-Kow}), (\ref{Es1,2}) is a
quadratic polynomial with respect to any of its three arguments
$z_1,\, z_2,\, s$. Its partial derivatives with respect to one of
variables are discriminants of the corresponding quadratic
equations. Squares of its partial derivatives with respect to one of
variables
 are factorized into functions of
 the rest two ones
 \bq
 \label{bi-factor}
\left(\frac{\p {\cal E}}{\p s}\right)^2= R(z_1,z_1)R(z_2,z_2),
\qquad \left(\frac{\p {\cal E}}{\p z_k}\right)^2= R(z_k,z_k)P_3(s),
\quad k=1,2.
 \eq
Here polynomial $P_3(s)$ is given by \bq \label{Eul-P3}
P_3(s)=4s^3-8H\,s^2+ 4H^2\,s-K\,s+ 4c^2A\,s+4c^2\,B
\eq
Because of complete differential of $\mathcal E(s,z_1,z_2)$
(\ref{Q-Kow}) is zero
\[
\frac{\p {\cal E}}{\p s}ds+\frac{\p {\cal E}}{\p z_1}dz_1+\frac{\p
{\cal E}}{\p z_2}dz_2=0\,,
\]
one gets relations between the differentials of the variables of
both types
\ben
\label{s-x-der} \frac{d s_{1,2}}{\sqrt{P_3(s_{1,2})}}
 =\frac{d z_1}{\sqrt{R({z_1,z_1})}} \pm  \frac{d z_2}{\sqrt{R({z_2,z_2})}}\,,
\en
which are the Euler equations (\ref{Eul-We}).

In matrix form the relations for velocities look like
\bq
\label{zt-st} \left(\begin{array}{c}
\dfrac{\dot {s}_1}{\sqrt{\varphi_1}} \\[2mm]
\dfrac{\dot {s}_2}{\sqrt{\varphi_2}}
\end{array}\right)=
\Psi \left(\begin{array}{c}
 {\dot {z}_1} \\[6mm]
 {\dot {z}_2}
\end{array}\right)\,,
\eq
where we denoted for brevity
\bq
\label{vphi1} \varphi_k\equiv P_3(s_k).
\eq
Signs at square roots in (\ref{s-x-der}), (\ref{zt-st})  are
compatible with definition of $s_1, s_2$ (\ref{s1,2}) and $\Psi$
(\ref{Psi}).

Using (\ref{R12}) -- (\ref{zt-st}) we can express integrals of
motion $H$ and $K$ (\ref{HKz}) in terms of cubic polynomial
$P_3(s)$, variables $s_{1,2}$ and their velocities $\dot{s}_{1,2}$
\ben \label{Hs} H&=&
\frac{s_1-s_2}2\,\left(\frac{\dot{s}_1^2}{\varphi_1}
-\frac{\dot{s}_2^2}{\varphi_2}\right)-\frac{s_1+s_2}2\,,\\
\label{Ks}
\frac{K}4&=&(2H+s_1+s_2)\lambda^2-\lambda\sqrt{-\varphi_1\varphi_2}
\left(\dsize\frac{\dot{s}_1}{\dsize\varphi_1}+\frac{\dot{s}_2}{\varphi_2}\right)\nn\\
&&+(s_1-s_2)\left(\frac{s_2\dot{s}_1^2}{\varphi_1}
-\frac{s_1\dot{s}_2^2}{\varphi_2}\right)-s_1s_2+H^2\,.
\en
Here the Hamiltonian $H$ and coefficients of integral $K$ at even
powers of gyrostatic parameter $\lambda$ are easy calculated using
definitions (\ref{s1,2}) and (\ref{zt-st}) only. For linear in
$\lambda$ term in $K=K_2\lambda^2+K_1\lambda +K_0$ one gets at first
\ben K_1=&-4i\frac{\dsize\dot s_1} {\dsize\sqrt{\varphi_1}} \left(
\sqrt{R(z_1, z_1)}\frac{\dsize\p }{\dsize\p z_1} -\sqrt{R(z_2,
z_2)}\frac{\dsize\p }{\dsize\p
z_2}\right)&\frac{R(z_1, z_2)}{(z_1-z_2)^2}\nn\\
\nn\\
&-4i\frac{\dsize\dot s_2} {\dsize\sqrt{\varphi_2}} \left(
\sqrt{R(z_1, z_1)}\frac{\dsize\p }{\dsize\p z_1} +\sqrt{R(z_2,
z_2)}\frac{\dsize\p }{\dsize\p z_2}\right)&\frac{R(z_1,
z_2)}{(z_1-z_2)^2}.\nn \en Due to inverse of (\ref{s-x-der}) one
converts derivatives $\p / \p z_{1,2}$ to $\p / \p s_{1,2}$
\[
K_1=-4i\left({\dot
s}_1\frac{\sqrt{\varphi_2}}{\sqrt{\varphi_1}}\frac{\p}{\p s_1}
+\dot{ s}_2\frac{\sqrt{\varphi_1}}{\sqrt{\varphi_2}}\frac{\p}{\p
s_2} \right)\frac{R(z_1, z_2)}{(z_1-z_2)^2}
\]
Minding that from definition (\ref{s1,2}) one gets $s_1+s_2=R(z_1,
z_2)/(z_1-z_2)^2 $ and including $i$ into square root we obtain finally
\bq
K_1= -4\left( \dot s_1 \sqrt{\frac{-\varphi_2}{\varphi_1}}
 +\dot s_2 \sqrt{\frac{-\varphi_1}{\varphi_2} }\right)\,.
\eq
In the section \ref{sect-ell} we recover these expressions of $H$
(\ref{Hs}) and $K$ (\ref{Ks}) using relation of the Kowalevski
gyrostat with the Clebsch system.

Equations (\ref{Hs}), (\ref{Ks}) have the form
\[\left.\Phi_{1,2}(s_1,s_2,\dot{s}_1,\dot{s}_2,A,B,H,K)\right|_\Sigma=0\,\]
and depend on the commuting variables $s_{1,2}$, their velocities
$\dot{s}_{1,2}$ and integrals of motion only. Excluding one of the
velocities we obtain two equations of fourth order in $\dot{s}_k$
\begin{equation}\label{s-GK}
\Bigl((s_1-s_2)^2\dot{s}_k^2+\lambda\sqrt{-\varphi_1\varphi_2}\dot{s}_k-
\beta_k\varphi_k\Bigr)^2+\lambda^2\left(\dot{s}_k^2
+\frac{(2H+s_1+s_2)\,\varphi_k}{s_1-s_2}\right)\varphi_k^2=0,
\end{equation} where
$k=1,2$ and $\beta_k$ is given by
\bq
\beta_k=(2H+s_1+s_2)\lambda^2+s^2+2Hs+H^2-\frac{K}{4}\,.
\label{beta_k}
\eq
At $\lambda=0$ the equations (\ref{s-GK}) are reduced to the
 Kowalevski top equations \cite{kow89,kotter}
\bq
\label{Kow-eq} (-1)^k\,(s_1-s_2)\dot{s}_k=\sqrt{P_5(s_k)\,}\,,\qquad
k=1,2,
\eq
which admit integration on $\Sigma$ (\ref{Int-Surf}) by Jacobi
inversion theorem. Here $P_5(s)=P_3(s)P_2(s)$ is a fifth order
polynomial, $P_3(s)$ is from (\ref{Eul-P3}) and
\bq\label{Kow-pol}
P_2(s)=s^2+2Hs+H^2-\frac{K}{4}
\eq
is a limiting value of $\beta_k$ (\ref{beta_k}),
$P_2(s_k)=\beta_k|_{\lambda=0}$.

To construct separation of variables for the gyrostat one needs to
substitute Lagrangian variables $\dot s_1, \dot s_2$ by momenta
$\pi_1 ,\pi_2$ conjugated to $s_1, s_2$ and to express integrals of
motion $H, K$ as functions of Hamiltonian variables. For the
Kowalevski top, i.e. at $\lambda=0$, momenta $\pi_1 ,\pi_2$ were
extracted from evolution equations (\ref{Kow-eq}) in \cite{ves2} and
\cite{komkuz1}. Integrals of motion as functions of $\dot s_1, \dot
s_2$, $\pi_1 ,\pi_2$ give rise to separation of variables with
separated equations of the form
\bq
s_i^2-4H_{top}s_i-\frac{4c^2 b^2}{s_i}+\vk_{top}=
2c^2(a^2-\frac{2b^2}{s_i})\cos{(2\sqrt{2s_i}\,\pi_i)}, \qquad i=1,2
\, , \label{top-SE}
\eq
where $\vk_{top} =4H_{top}^2-K_{top}+2{c^2}{a^2}$.

At $\lambda\neq 0$ transition from velocities $\dot s_1, \dot s_2$
to the corresponding momenta is unknown due to complicated form of
equations of motion (\ref{s-GK}) in $s_{1,2}$ variables, thus we
cannot claim that $s_{1,2}$ are separation variables. Nevertheless,
below we prove that equations (\ref{s-GK}) may be solved in
quadratures using relation of the Kowalevski gyrostat with the
Clebsch system.

\section{The Kowalevski gyrostat and the Clebsch system}
\label{sect-KC} \setcounter{equation}{0}

Let two vectors $\bfit l$ and $\bfit p$ are coordinates on the phase
space $\mathcal M$. As a Poisson manifold $\mathcal M$ is identified
with the algebra $e(3)^*$ equipped with brackets
\bq\label{new-e3} \{l_{i}, l_{j}\}=\varepsilon_{ijk}l_{k}, \quad
\{l_{i}, p_{j}\} =\varepsilon_{ijk}p_{k}, \quad \{p_{i},
p_{j}\}=0\,. \eq These brackets respect two Casimir elements
\bq\label{caz-Cl} \mathcal A=(\bfit p,\bfit p),\qquad \mathcal
B=(\bfit p,\bfit l)\,. \eq
 The following integrable case for the Kirchhoff equations on
$e(3)$ was found by Clebsch \cite{clebsch}
\bq \label{Cl-eqm}
\mathcal X\left(%
\begin{array}{cc}
 \bfit l \\
 \bfit p
\end{array}%
\right) : \qquad\dot {\bfit l}=\bfit p\times \Q\,\bfit p\,, \qquad
\dot {\bfit p}= \bfit p \times \bfit l\,.
\eq
Here $\Q$ is a constant symmetric matrix, $\det \Q\neq 0\,$.
Equations of motion (\ref{Cl-eqm}) are generated  by the brackets
(\ref{new-e3}) and the Hamilton function
\bq
\label{H-cl}
\mathcal H=\frac12\,{\bfit l}^2+\frac12\,(\Q\, {\bfit p},{\bfit p}).
\eq
The second integral of motion reads as
\bq\label{K-cl} \mathcal
K=(\Q\,{\bfit l},{\bfit l})- (\Q^{\vee}{\bfit p},{\bfit p}),
\eq
where $\Q^{\vee}$ stands for adjoint matrix, i.e. cofactor matrix.
In our case it reads $\Q^{\vee}=({\det \Q})\,\Q^{-1}$.

The vector field $\mathcal X$ (\ref{Cl-eqm}) is bi-hamiltonian
vector field
\bq\label{bi-Cl}
\mathcal X\left(
\begin{array}{c}
 \bfit l \\
 \bfit p
\end{array}
\right)=\mathcal P_0d\mathcal H=\mathcal P_1d\mathcal K\,,
\eq
where
\[
\mathcal P_0=\left(%
\begin{array}{cc}
 \mathbf L & \mathbf P \\
 -\mathbf P & 0
\end{array}%
\right),\qquad
\mathcal P_1=\frac12\left(%
\begin{array}{cc}
 \Q^{-1} & 0 \\
 0 &\mathbf I \\
\end{array}%
\right)\mathcal P_0\left(%
\begin{array}{cc}
 \Q^{-1} & 0 \\
 0 & \mathbf I \\
\end{array}%
\right),
\]
and
\bq \label{L-lax}
\mathbf L=\left(%
\begin{array}{ccc}
  0& l_3 & -l_2\\
  -l_3& 0 & l_1\\
  l_2& -l_1 & 0
\end{array}%
\right)\,,\qquad
\mathbf P=\left(%
\begin{array}{ccc}
  0 & p_3 & -p_2 \\
  -p_3 & 0 & p_1 \\
  p_2 & -p_1 & 0
\end{array}%
\right)\,.
\eq
Here $\mathbf I$ stands for $3\times 3$ unit matrix. Poisson
matrices $\mathcal P_0$ and $\mathcal P_1$ define two compatible
linear brackets on $\mathcal M$ in a standard bi-hamiltonian
formulation.
%The similar family of brackets $\mathcal P_0+\mu\mathcal P_1$ was
%introduced for the Euler top in \cite{hm91}.

\begin{rem}\label{rem-Neum} If $\B=0$ the flow (\ref{Cl-eqm}) is equivalent
to that of the Neumann system with the Newton equation of motion
\bq\label{Neum-eqm}
\ddot{\bfit p}=-\Q\bfit p+\Bigl( (\Q\bfit p,\bfit p)-\dot{\bfit
p}^2\Bigr)\bfit p
\eq
describing the motion of a mass point on the sphere $\bfit p^2=\A$
under influence of the force $-\Q\bfit p$.
\end{rem}

\subsection{Mapping of the Kowalevski gyrostat flow onto the Clebsch flow}

The idea of the map Kowalevski top flow (\ref{Kow-eqm}) onto the
Neumann flow (\ref{Neum-eqm}) originally appeared in Heine and
Horosov \cite{HH,P02} to the Kowalevski top and was extended to
$so(4), so(3,1)$ in \cite{komkuz}.

Let us introduce the following complex vector-functions
\bq\label{p}
 {\bfit p}=\alpha\left(-i\frac{J_1}{J_2},\,
\frac{J_1^2+J_2^2+1}{2J_2},\,
i\frac{J_1^2+J_2^2-1}{2J_2}\right),\qquad \alpha\in\mathbb C,
\eq
and
\bq \label{l} {\bfit
l}_{\mbox{\fs
top}}=\left(-i\frac{cx_3}{J_2},\,\frac{2cx_3J_1-J_3(J_1^2+J_2^2-1)}{2J_2},\,
i\frac{2cx_3J_1-J_3(J_1^2+J_2^2+1)}{2J_2}\right),
\eq
such that
\bq
\label{lp-t} \A=({\bfit p},\,{\bfit p})=\alpha^2, \qquad
\B=({\bfit p},\,{\bfit l}_{\mbox{\fs top}})=0.
\eq
We permuted the first and the second entries in original vectors
\cite{HH} to make gyrostat formulas slightly more symmetric.

In order to describe mapping of the gyrostat flow (\ref{Kow-eqm})
onto the Clebsch flow (\ref{Cl-eqm}) we have to shift vector
$\bfit l_{\mbox{\fs top}}$ by the rule \bq\label{lg} \bfit l=\bfit
l_{\mbox{\fs top}}+\alpha^{-1}\lambda\,\Bigl(\bfit p+i\,\bfit
k\times \bfit p\Bigr)\,,\qquad \eq where $\bfit k=(1,0,0)$ is a
unit vector. In compare with (\ref{lp-t}) scalar product of vectors
${\bfit l}$ and ${\bfit p}$ for gyrostat becomes differ from zero
\bq \label{lp-g} \B=({\bfit p},\,{\bfit l})=\alpha\lambda\,. \eq

Adding constraints $A=a, B=b$ to relations (\ref{p}), (\ref{lg}) one
gets correspondence $M\simeq \mathcal M$ of the phase manifolds for
the Kowalevski gyrostat and  the Clebsch system. Initial Poisson
structure on $M$ (\ref{e3}) gives rise to the cubic Poisson brackets
$\{\, \cdot , \cdot\}_3$ on $\mathcal M$, for instance
\[
\{p_i,p_k\}_3\equiv\{p_i(x,J),p_k(x,J)\}
=\varepsilon_{ijk}\,p_k\,(l_2+il_3)(p_2+ip_3)\,.
\]
We shall not use these induced brackets directly and, therefore,
remaining brackets are omitted.

On the other hand linear Poisson structure on $\mathcal M$
(\ref{new-e3}) gives rise to the cubic Poisson brackets $\{\, \cdot
, \cdot\}_3$ on $M$. So, $M$ and $\mathcal M$ are multi-Poisson
manifolds for which we constructed the correspondence
$M\simeq\mathcal M$ such that their linear brackets map to cubic
brackets and vise versa.

\begin{rem} For the Kowalevski top and gyrostat variables
$\bfit p, \bfit l_{\mbox{\fs top}}$ and $\bfit p, \bfit l$ are
coordinates on the different spaces $\mathcal M_{\mbox{\fs top}}$
and $\mathcal M$ with different brackets (\ref{new-e3}) $\{\cdot ,
\cdot \}_{\mbox{\fs top}}$ and $\{\cdot , \cdot \}$ forming two
sample of $e(3)$ algebra (\ref{new-e3}). With respect to the top
brackets $\{\cdot , \cdot \}_{\mbox{\fs top}}$ the gyrostat
variables $\bfit p, \bfit l$ form the central extension of
$e(3)_{\mbox{\fs top}}$ which is contracted to $e(3)_{\mbox{\fs
top}}$ in the limit $\lambda \to 0 $.
\end{rem}

Similar to the top \cite{HH} let us introduce symmetric matrix $\Q$
depending on integrals of motion of the Kowalevski gyrostat and the
Casimir elements on the initial algebra (\ref{e3})
\bq
\label{Q-matrix} \Q = \alpha^{-2}\left(
\begin{array}{rcl}
-H &\quad   -i c b      &\quad i c b\\[3mm]
-i c b  &\quad -\frac14 +c^2 \vk &\quad i\left(\frac14 +c^2 \vk\right) \\[3mm]
i c b &\quad i\left(\frac14 +c^2 \vk\right) &\quad \frac14 -c^2 \vk
\end{array} \right) ~,\qquad\vk=a- {K}/{4c^2}.
\eq
It is easy to prove that this matrix remains constant with respect
to dynamics of the Clebsch system on $\mathcal M$ generated linear
and cubic Poisson structures.

\begin{prop}\label{main-Th}
Let us identify $\mathcal M$ with $M$ by the map $\{\bfit x,\bfit
J\}\to \{\bfit p,\bfit l\}$ (\ref{p}), (\ref{lg}) such that the
Casimir elements are equal to
\bq\label{C-Cl}
 A=a,\qquad B=b,\qquad \A=\alpha^2,\qquad \B=\alpha\lambda\,,
\eq
If the matrix $\Q$ is given by (\ref{Q-matrix}), then
\bq\label{HK-HK}
2\mathcal H=-H+\lambda^2\,,\qquad 4\alpha^2\,\mathcal
K=K-4\lambda^2\,H.
\eq
and vector field $X$ (\ref{Kow-eqm}) for the Kowalevski gyrostat on
$M$ coincides with vector field $\mathcal X$ (\ref{Cl-eqm}) for the
Clebsch system on $\mathcal M$
\bq\label{m-eq}
X=P_0dH=\mathcal P_0\mathcal H=\mathcal X.
\eq
The similar equality holds for the second commuting flows of the
Kowalevski gyrostat and the Clebsch system.
\end{prop}
The proof is straightforward.

 According to (\ref{m-eq}) on the space $\mathcal
M$ initial linear Poisson structure $\mathcal P_0$ and cubic Poisson
structure induced by $P_0$ generate the same vector field
$X=\mathcal X$ with respect to a common integral $\mathcal H\simeq
H$ (\ref{m-eq}). We can embed (\ref{m-eq}) in a standard
bi-hamiltonian formulation with two functionally different integrals
of motion $\mathcal H$ and $\mathcal K$ (see (\ref{bi-Cl})), if we
extend Poisson space $\mathcal M\simeq M$ by additional degree of
freedom considering $\lambda$ as an independent dynamical variable.
Similar extension was used by Sklyanin \cite{S-GC} when he
constructed Lax matrix for the quantum Goryachev-Chaplygin gyrostat.

Thus we arrive at one of the main results of the paper:
\begin{prop}
Solutions of the Clebsch problem give rise to solutions of the
Kowalevski gyrostat and vise versa.
\end{prop}
We can get solution of the Kowalevski gyrostat problem using either
the Kobb-Kharlamova quadratures \cite{mink,kobb,harl}, or the
K\"otter solution \cite{kot92} of the Clebsch system in theta
functions.

Recall once more that for $\lambda=0$ solutions of the Neumann
system was identified with the Kowalevski solutions of her problem
in \cite{HH,komkuz}.

\section{ Lax representations}
\setcounter{equation}{0}

Equations of motion for the Clebsch system may be expressed in a Lax
form
\bq\label{Lax-g}
\frac{d}{dt}{\mathscr L}(y)=\left[{\mathscr L}(y),{\mathscr
A}(y)\right],
\eq
which automatically exhibits  constants of motion as eigenvalues of
$\mathscr L$ and leads to the linearization of the flow on the
Jacobi or Prym varieties of the algebraic curve $\det\left( \mathscr
L(y)-\mu\mathbf I\right)=0$. Here $y$ is an auxiliary variable
(spectral parameter).

There are few different Lax matrices for the Clebsch system
associated with two different integration procedures. These matrices
depend on rational and elliptic matrix-functions of the spectral
parameter.

The rational $3\times 3$ Lax matrix for the Clebsch system was found
by Perelomov \cite{P81}
\bq\label{Lax-33}
{\mathscr L}_r(y)=\Q +\mathbf L y - \mathbf N y^2\,,\qquad {\mathscr
A}_r(y)=y^{-1}\Q+\mathbf L\,,
\eq
where $\Q$ is symmetric matrix (\ref{Q-matrix}), $\mathbf L$ is
given by (\ref{L-lax}) and $\mathbf N={\bfit p}\otimes {\bfit p},
\quad \mathbf N_{ij}=p_{i}p_{j}$. The corresponding spectral curve
is equal to
\bq\label{tau}
\tau_1(y, \mu)=\B^2y^4+\Bigl(\A\mu^2+(2\H-\A\mbox{\rm tr}\,
\Q)\mu-\K\Bigr)y^2-\det\,(\Q-\mu\mathbf I)=0\,.
\eq
We do not know separated variables for the Clebsch system associated
with this curve. If $\mathcal B=0$ the separated variables
are well-known \cite{neum59} and may be
obtained by various integration schemes, for instance, by
intersection of two algebraic curves related with Lax matrix
${\mathscr L}_r(y)$ \cite{komkuz}. The regular way is provided by
the Sklyanin method \cite{skl95} that is appeared to be nontrivial
in the considered case:

\begin{prop} In the Neumann case at $\mathcal B=0$ the separated variables
$u_{1,2}$ are poles of the corresponding Baker-Axiezer function
$\boldsymbol \Psi$, such that
\[{\mathscr
L}_r(y)\boldsymbol \Psi=\mu\boldsymbol \Psi,\qquad\mbox{\rm
and}\qquad (\boldsymbol \alpha, \boldsymbol \Psi)=1,
\]
with dynamical nor\-ma\-li\-za\-ti\-on $\boldsymbol \alpha={\bfit
p}\equiv({p}_1,{p}_2,{p}_3)$. Canonical variables $u_k$ and their
momenta $p_{u_k}$ lie on the algebraic curve (\ref{tau}) that gives
rise to separated equations.
\end{prop}

The proof consists of direct comparison of the known separated
variables \cite{neum59} with poles of the Baker-Axiezer functions.
Integration procedure of the generic Clebsch system in these
variables will be considered in the next section.

Another Lax matrices for the Clebsch system is related with
K\"otter's approach \cite{kot92}. Let  $\Q=\mbox{\rm
diag}\,(a_1,a_2,a_3)$ is a diagonal matrix. Introduce two vectors
$\bfit t(\mu)$ and $\bfit s(\mu)$ \bq\label{ts-vect} \bfit t=\mathbf
W(\mu)\,\bfit l+\mathbf W\,^{\vee}(\mu)\,\bfit p,\qquad \bfit
s=\mathbf W(\mu)\,\bfit p,\qquad \mathbf W=\bigl(\mu\mathbf
I-\Q\bigr)^{1/2}\,, \eq where $\mathbf W(\mu)=\mbox{\rm
diag}(w_1,w_2,w_3)$ is diagonal
 and $\mathbf W\,^{\vee}$ is its adjoint matrix.
In a special uniformisation of the spectral parameter $\mu$ diagonal
entries $w_k=\sqrt{\mu-a_k}$  can be considered as basic elliptic
functions (see \cite{kot92,bbe94,skl98}).

\par\noindent
The equations of motion
\bq\label{st-eqm}
\dot{\bfit t}(\mu)=\bfit s(\mu)\times \bfit t(\mu)\,,
\eq
may be rewritten in the Lax form (see \cite{bbe94}) using matrix
\bq\label{Lax-K}
{\mathscr L}_e(\mu)=\sum_{k=1}^3t_k(\mu)\sigma_k\equiv
\sum_{k=1}^3\left(w_k l_k+\frac{w_1w_2w_3}{w_k}p_k \right)\sigma_k
\,,\qquad{\mathscr A}_e(\mu)=\sum_{k=1}^3s_k(\mu)\sigma_k\,,
\eq
where $\sigma_k$ are the Pauli matrices.

The corresponding spectral curve reads
\bq\label{tau2}
\tau_2(w,\mu)=w^2-\bfit
t^2(\mu)=w^2-\Bigl(\A\mu^2+(2\H-\A\,\mbox{\rm
tr}\,\Q)\mu-\K\Bigr)-2\B\sqrt{-\det\,(\Q-\mu\mathbf I)}=0\,.
\eq
The linearization of the flow associated with the curve (\ref{tau2})
and expressions of initial variables $\bfit l$ and $\bfit p$ in
theta-functions were done by K\"otter \cite{kot92} using new
Lagrangian variables $\mathpzc z_{1,2},\dot{\mathpzc z}_{1,2}$ which
satisfy to  nice evolutionary equations and may be considered as
candidates for separation variables. These variables will be studied
in the next section.

According to \cite{bob83} let us consider one parametric
transformation $f_\mu:so(4)\to e(3)$
\bq \label{map-Cl}
f_\mu:\qquad p_i=w_i(S_i-T_i)\,,\qquad
l_i=\dfrac{w_1w_2w_3}{w_i}(S_i+T_i)\,.
\eq
where $S_i,T_i$ are coordinates on $so(4)=so(3)\oplus so(3)$ with
the Lie-Poisson brackets
\begin{equation} \label{2o3}
\bigl\{ S_i\,,S_j\,\bigr\}= \varepsilon_{ijk}\,S_k\,, \qquad \bigl\{
S_i\,,T_j\,\bigr\}= 0\,,\qquad \bigl\{
T_i\,,T_j\,\bigr\}=\varepsilon_{ijk} T_k\,.
\end{equation}

\begin{rem}\label{kott-bob}
The inverse transformation $f_\mu^{-1}$ reads as
\bq
\label{ST}
S_i=\dfrac{w_i}{2w_1w_2w_3}\,l_i+\dfrac{1}{2w_i}\,p_i,\qquad
T_i=\dfrac{w_i}{2w_1w_2w_3}\,l_i-\dfrac{1}{2w_i}\,p_i\,,
\eq
where $w_k$ depends on the parameter $\mu$. The map $f_\mu^{-1}$ is
easy generalized to two parametric mapping $f_{\mu,\nu}^{-1}$ if we
substitute $w_k(\mu)=\sqrt{\mu-a_k}$ and $w_k(\nu)=\sqrt{\nu-a_k}$
in the definition $S_k$ and $T_k$ respectively.

In Section 7 we show that K\"otter used namely this mapping to
integrate the Clebsch system.
\end{rem}

The mapping $f_\mu$ is a twisted Poisson map, which identifies two
bi-Hamiltonian manifolds $ e(3)$ and $ so(4)$ such that
\[\begin{array}{c}
e(3)\qquad\,\,\, so(4)\\
\xymatrix{ \{.\,,.\} \ar[dr] & \ar[dl]\{.\,,.\}
\\
\{.\,,,\}^* \ar[ur] & \ar[ul]\{.\,,.\}^*}
\end{array}\qquad\mbox{\rm instead of}\qquad
\begin{array}{c}
e(3)\qquad\,\,\, so(4)\\
\xymatrix{ \{.\,,.\} \ar[r] & \ar[l]\{.\,,.\}
\\
\{.\,,.\}^* \ar[r] & \ar[l]\{.\,,.\}^* }\end{array}\quad\mbox{\rm
for the usual Poisson map}.
\]
Here second compatible brackets $\{.\,,.\}^*$ on $e(3)$ are equal to
\bq\label{sec-e3} \{l_{i}, l_{j}\}^*=\varepsilon_{ijk}w_k^2l_{k},
\quad \{l_{i}, p_{j}\}^* =\varepsilon_{ijk}w_jp_{k}, \quad \{p_{i},
p_{j}\}^*=\varepsilon_{ijk}l_{k}\,. \eq The polynomial $\bfit
t^2(\nu)$ is a Casimir function of the corresponding Poisson pencil
$\{.\,,.\}_\nu=\{.\,,.\}-\nu\{.\,,.\}^*$, where $\nu\neq \mu$. For
brevity the second linear brackets $\{.\,,.\}^*$ on $so(4)$ will be
omitted because they are completely determined by mapping $f_\mu$
(\ref{map-Cl}). The similar twisted Poisson map related to Steklov
integrable cases on $e(3)$ and $so(4)$ is discussed in \cite{ts04b}.

The  map $f_\mu$ identifies the Clebsch system on $e(3)$ with
the Schottky--Manakov system on $so(4)$ and, according to
\cite{bob83}, identify the corresponding $2\times 2$ elliptic Lax
matrices. At the same time the Schottky--Manakov system on $so(4)$ has
another $4\times 4$ Lax matrix with linear dependence on spectral
parameter. It allows us to construct two-parametric family of Lax
matrices for the Clebsch system
\bq\label{Lax-munu}
\widetilde{{\mathscr L}}(\nu,\mu)=
\left(%
\begin{array}{cc}
  \nu\mathbf W^\vee+\mathbf W^\vee\,\mathbf L\,\mathbf W^{-1} &\mathbf W^\vee \bfit p\\
 \\
 -(\mathbf W^\vee \bfit p)^T& 0
\end{array}%
\right)\,,\qquad \widetilde{{\mathscr A}}(\nu,\mu)=
\left(%
\begin{array}{cc}
 0 &\mathbf W \bfit p\\
 \\
 -(\mathbf W \bfit p)^T& \nu\det\mathbf W
\end{array}%
\right)\,,
\eq
where $\mathbf W=\bigl(\mu\,\mathbf I-\Q\bigr)^{1/2}$ and $\mathbf
L$ is given by (\ref{L-lax}). This family of Lax matrices leads to a
spectral surface, rather than a spectral curve. The nature of this
surface is discussed in \cite{avm88}.

At $\mu=0$ the Lax matrix (\ref{Lax-munu}) becomes rational
 with the following spectral
curve
\ben
\tau_3(y,\nu)&=&y^4-\mbox{\rm tr}\,\Q^\vee\,\nu
y^3+\left(\det\Q\,\mbox{\rm
tr}\,\Q\,\nu^2-\mathcal K\right)y^2\nn\\
\label{tau-3}\\
&-&\det\Q\Bigl(\nu^2\det\Q-2\mathcal H+\mbox{\rm tr}\,\Q\,\mathcal
A\Bigr)\nu\,y-\det\Q\Bigl(\mathcal A\,\det\Q\,\nu^2+\mathcal
B^2\Bigr)=0\,.\nn \en

The separated variables for the Clebsch system associated with this
curve are unknown. The corresponding solutions in theta-functions
were obtained  by algebro-geometric methods in \cite{zhiv98}.

At $\nu=0$ the Lax matrix (\ref{Lax-munu}) becomes elliptic matrix,
which spectral curve coincides with spectral curve (\ref{tau}) of
the rational Lax matrix up to transformation $\widetilde{y}=\mathcal
B\,y$.

\subsection{Algebraic curves associated with the Kowalevski gyrostat}
From relation of the Kowalevski gyrostat with the Clebsch system
established in the section (\ref{sect-KC})
 one gets three Lax matrices
$\mathscr L_r(y)$, $\mathscr L_e(\mu)$ and $\widetilde{\mathscr
L}(\mu)$ for the Kowalevski gyrostat.

The spectral curve of the rational Lax matrix $\mathscr L_r(y)$
looks like
\bq\label{tau-Kow}
\tau_1(y,\mu)=\lambda^2\,y^4+\left(\mu^2+(\mu+H)\lambda^2-\frac{K}4\right)y^2-\frac{P_3(\mu)}4\,,
\eq
where $P_3(\mu)$ is given by (\ref{Kow-pol}). It is a biquadratic
function of $y$ with coefficients being quadratic and cubic
polynomials of $\mu$. At $\lambda=0$ it is a famous Kowalevski curve
and the corresponding variables $u_{1,2}$ give rise to the
Kowalevski separated variables $s_{1,2}$ (see discussion in the next
section).

The $2\times 2$ elliptic matrix ${\mathscr L}_e(\mu)$ for the
Kowalevski gyrostat has another spectral curve
\[
\tau_2(w,\mu)=w^2-\left(\mu^2+(\mu+H)\lambda^2-\frac{K}4\right)-\lambda\sqrt{P_3(\mu)}\,.
\]
Associated with this curve expressions of initial variables $\bfit
x$ and $\bfit J$ in theta-functions may be obtained using K\"otter
formulae and results of the section (\ref{sect-KC}).

At $\lambda=0$ the two-dimensional family of $4\times 4$ Lax matrix
$\widetilde{{\mathscr L}}(\nu,\mu)$ (\ref{Lax-munu}) was obtained by
Adler and van Moerbeke \cite{avm88} using directly
correspondence of the Kowalevski top flow and the Schottky-Manakov
flow. For the Kowalevski gyrostat third algebraic curve
(\ref{tau-3}) is given by
\ben
\tau_3(y,\nu)&=&y^4-\left(Ac^2-\frac{K}4\right)\,\nu\,y^3 +\left(
\nu^2\left(Ac^2-\frac{K}4\right)H^2-(cB\nu-\lambda)(cB\nu+\lambda)H-\frac{K}{4}
\right)\,y^2\nn\\
\nn\\
 &-&\left(\left(Ac^2-\frac{K}4\right)H-c^2B^2\right)
\left(\nu^2\left(Ac^2-\frac{K}4\right)H-(cB\nu-\lambda)(cB\nu+\lambda)\right)(y\nu-1)\nn
\en
Associated with this curve solutions of the gyrostat may be obtained
using theta-functions expressions for the Clebsch variables from
\cite{zhiv98} and change of variables from the section 3.

The fourth algebraic curve associated with gyrostat is due to
rational Lax matrices of \cite{brs89}. It is given by
\[\tau_4(y,\mu)=y^4-y^2\left(2(\mu^2-H)-\lambda^2+\frac{c^2A}{\mu^2}\right)
+(\mu^2-H)^2-\frac{K}{4}+c^2A-\frac{c^2B^2}{\mu^2}\,.
\]

\section{Integration of the Clebsch system in elliptic coordinates}
\setcounter{equation}{0} \label{sect-ell}

 Minkowski \cite{mink} identified the Clebsch system
with the Jacobi problem of geodesic motion on ellipsoid for which
elliptic coordinates $u_{1,2}$ were introduced by Jacobi. In 1895
Kobb started the integration procedure in the Euler angles and
passed to variables $\xi=\tan(\theta/2)$, $\nu=\tan(\phi/2)$, which
are equivalent to variables $u_{1,2}$ \cite{kobb}. In 1959
Kharlamova \cite{harl} used directly elliptic coordinates $u_{1,2}$ for
integration of the second flow of the Clebsch system associated with
$\mathcal K$.

In order to explain the method proposed in \cite{harl,kobb} we
reproduce some simple formulae. Using equations of motion
(\ref{Cl-eqm}) and the Casimir elements (\ref{caz-Cl}) we express
angular momenta $\bfit l$ via
 Lagrangian variables ${\bfit p}, \dot{\bfit p}$
\[
{\bfit {l}}=\frac{1}{\A}\left(\B\bfit p+ \dot{\bfit p}\times\bfit
p\right)\,.
\]

Then we introduce variables $u_{1,2}$ as roots of the following
function
\bq\label{seq-Cl}
e(\mu)=(\mu-u_1)(\mu-u_2)= \mu^2+\left(\frac{(\Q\bfit p,\bfit
p)}{\A} -\mbox{\rm tr}\,\Q\right)\mu+\frac{(\Q^{\vee}{\bfit
p},{\bfit p})}{\A}\,.
\eq
Substituting $u_{1,2}$ and their velocities $\dot{u}_{1,2}$ into
 (\ref{H-cl}) and (\ref{K-cl}) one gets the
Hamilton function $ \mathcal H=\mathcal T+\mathcal V$
\ben\label{Hu-Cl}
\mathcal T&=&
\frac{u_1-u_2}{2}\left(\frac{\dot{u_1}^2}{\varphi_1}-\frac{\dot{u_2}^2}{\varphi_2}\right)
+\frac{\B^2}{2\A}\,,\qquad \mathcal V=\frac12\Bigl(\mbox{\rm
tr}\,\Q-u_1-u_2\Bigr)\, \A,
\en
and the second integral of motion
\[
\mathcal K =(u_1-u_2)
\left(\frac{u_2\,\dot{u_1}^2}{\varphi_1}-\frac{u_1\,\dot{u_2}^2}{\varphi_2}\right)
-\frac{\B}{\sqrt{\A}}\left(\dot{u_1}\sqrt{-\frac{\varphi_2}{\varphi_1}}
+\dot{u_2}\sqrt{-\frac{\varphi_1}{\varphi_2}}\right)
+\frac{\B^2}{\A}\Bigl(\mbox{\rm tr}\,\Q-u_1-u_2\Bigr) -u_1\,u_2\A\,.
\]
in terms of variables $u_{1,2}$, their velocities $\dot{u}_{1,2}$
and the cubic polynomial
\[\varphi_k=4\det\,(\Q-u_k\mathbf I)\,.\]
Below this polynomial will be identified with the cubic polynomial
$\varphi_k$ (\ref{vphi1}) for the Kowalevski gyrostat for which we
used the same notation.

Excluding one of the velocities from these equations we obtain two
equations of fourth degree in each of velocities depending on both
variables $u_1$ and $u_2$
\ben \label{cl-4eq} \Bigl( \mathcal
A(u_1-u_2)^2\dot{u}^2_k&+&\mathcal B\sqrt{-\mathcal
A\varphi_1\varphi_2}\dot{u_k}+\beta_k\varphi_k\Bigr)^2\\
\nn\\
&+&\mathcal B^2\left(\mathcal A\dot{u}_k^2+\frac{(\mathcal
A^2(u_1+u_2-\mbox{\rm tr}\,\Q)+2\mathcal A\mathcal H-\mathcal
B^2)\varphi_k}{u_1-u_2} \right)\varphi_k^2=0\,.\nn
\en
Here $\beta_k$ is a cubic polynomial also depending on $u_1$ and
$u_2$
\bq
\label{beta} \beta_k=\B^2(u_1+u_2+u_k-\mbox{\rm tr}\,\Q)+\A\Bigr(\A
u_k^2+(2\mathcal H -\A \mbox{\rm tr}\,\Q)\,u_k -\mathcal K)\,.
\eq
Momenta conjugated to ${u}_{1, 2}$ are introduced by
%\bq\label{mom-u}
$p_{u_{1,2}}=\frac{\partial {\mathcal T}}{\partial\dot{u}_k}$,
%\eq
where $\mathcal T$ is kinetic energy (\ref{Hu-Cl}) and the Casimir
operator $\B$ depends on velocities $\dot{u}_k$.

Equations (\ref{cl-4eq}) were solved in quadratures in
\cite{harl,kobb}. We have to underline only that variables
$\{u_{1,2},p_{u_{1,2}}\}$ are not the separated variables in the
standard meaning.

\begin{rem}
After a suitable rotation
\[\tilde{\bfit p}=V\bfit p,\qquad\tilde{\bfit l}=V\bfit
l,\qquad \Q\to \widetilde{\Q}=V\Q V^{-1}=\mbox{diag}(a_1,a_2,a_3)
\]
which diagonalize the matrix $\Q$, coordinates $u_{1,2}$ coincide
with elliptic coordinates on sphere ${\bfit p}^2={\mathcal A}$
defined by
\bq\label{ell-cl}
e(\mu)=(\mu-u_1)(\mu-u_2)= \frac{\det\,(\Q-\mu\mathbf I)}{\A}\left(
\frac{\tilde{p}_1^2}{\mu-a_1}
+\frac{\tilde{p}_2^2}{\mu-a_2}+\frac{\tilde{p}_3^2}{\mu-a_3}\right).
\eq
\end{rem}

\begin{rem} \label{rem-Neu}
If $\B=0$ there are considerably simple equations
\bq\label{sep-Neu}
\Bigl(\A(u_1-u_2)^2\dot{u}_k^2+4\varphi_k\beta_k\Bigr)^2=0\, \qquad
k=1,2.
\eq
In this case both integrals are quadratic polynomials in momenta $
p_{u_{1,2}}=\pm\frac{(u_1-u_2)\dot{u_k}}{2\varphi_k}$ and the
corresponding Neumann system belongs to the St\"ackel family of
integrable systems. Moreover, variables $\{u_k,p_k\}$ are separated
variables which lie on the spectral curve (\ref{tau}) of the
rational Lax matrix.
\end{rem}

\subsection{The Kowalevski gyrostat in $s$ variables}

Without loss of generality we put $\alpha=|\bfit p|=1$ . Inserting
$\bfit p$ (\ref{p}) and $\bfit l$ (\ref{lg}) into the generating
function $e(\mu)$ (\ref{seq-Cl}) of $u$-variables one gets
\[
e(\mu)=\frac{1}{(z_1-z_2)^2}{\cal E}(s){\mid}_{s={-\mu-H}}\,,
\]
where ${\cal E}(s)$ (\ref{Es1,2}) is the generating functions of the
$s$-variables. Combining this fact with the Proposition \ref{main-Th}
we have
\bq
\label{s-u} u_k=-s_k-H,\qquad \dot{u}_k=-\dot{s_k}\,.
\eq
Here $\dot{s_k}=\{H,s_k\}_1$ and $\dot{u_k}=\{\mathcal H,u_k\}_2$
and $\{\}_{1,2}$ means the Poisson brackets (\ref{e3}) on $ M$ and
the Poisson brackets (\ref{new-e3}) on $\mathcal M$, respectively.

Inserting constant of motion (\ref{HK-HK}), (\ref{C-Cl}) and
variables (\ref{s-u}) into the equations (\ref{Hu-Cl}) we arrive
at the same expressions of integrals $H$ and $K$
(\ref{Hs})-(\ref{Ks}) in terms of $s$-variables and polynomials
$\varphi_{1,2}$. However, in contrast with the Clebsch system
expression for Hamiltonian $H$ (\ref{Hs}) contains both integrals
$H$ and $K$ via polynomial $\varphi(s)$. Expressions of $H$ and
$K$ in $s_{1,2}$ and $\dot{s}_{1,2}$ variables may be obtained by
solving equations (\ref{s-GK}). Unfortunately, we cannot use very
complicated resulting formulae to calculate associated with $s$
variables momenta as in the Clebsch case.

\section{The K\"otter solution of the Clebsch system}
\setcounter{equation}{0}

In 1888 Minkowski proved that the Clebsch flow is isomorphic to
geodesic flow on the ellipsoid \cite{mink}. Then in 1891 Schottky
found that the Clebsch flow is isomorphic to integrable motion of
four-dimensional rigid body, which may be integrated in a special
case \cite{shott91} associated with the Neumann system. In 1892
K\"otter \cite{kot92} joined these results together and integrated
the Clebsch system completely.

Very brief description of K\"otter approach was presented in
\cite{dubr}. Below we reproduce the essence of K\"otter derivation
in modern notations. Instead of two vectors $\bfit t(\mu)$ and
$\bfit s(\mu)$ (\ref{ts-vect}) defining pair of the Lax matrices
${\mathscr L}_e(\mu)$ and ${\mathscr A}_e(\mu)$ (\ref{Lax-K}) we
take two vectors $\bfit t(\mu)$ and $\bfit t(\nu)$ defining two
samples of the first Lax matrix ${\mathscr L}_e$ and pass to their
linear combinations \bq\label{gen-xieta}
\begin{array}{c}
  { \xi}=a(\mu,\nu)\,\bfit t(\mu)+b(\mu,\nu)\,\bfit
  t(\nu)=\mathbf W_+(\mu,\nu)\bfit l+\widetilde{\mathbf W}_+(\mu,\nu)\bfit p\,,
  \\
  \\
   \eta=a(\mu,\nu)\,\bfit t(\mu)-b(\mu,\nu)\,\bfit t(\nu)=
   \mathbf W_-(\mu,\nu)\bfit l+\widetilde{\mathbf W}_-(\mu,\nu)\bfit
   p\,,
\end{array}
\eq
depending on two auxiliary functions $a,b$ having zero time
derivatives $\dot{a}=\dot{b}=0$. Here we denoted for brevity
\[\mathbf W_\pm(\mu,\nu)=a(\mu,\nu)\mathbf W(\mu)\pm b(\mu,\nu)\mathbf
W(\nu),\qquad \widetilde{\mathbf W}_\pm(\mu,\nu)=a(\mu,\nu)\mathbf
W(\mu)^\vee\pm b(\mu,\nu)\mathbf W\,^\vee(\nu)\,.
\]
Using inverse relations
\bq\label{inv-relK}
\bfit l=-\mathbf Z^{-1}\Bigl(\widetilde{\mathbf W}_-\,
\xi-\widetilde{\mathbf W}_{+}\, \eta\Bigr),\qquad \bfit p=\mathbf
Z^{-1}\Bigl(\mathbf W_{-}\, \xi-\mathbf W_{+}\, \eta\Bigr),
\eq
where
\[
\mathbf Z=2a(\mu,\nu)b(\mu,\nu)\Bigl( \mathbf W(\mu)\mathbf
W^\vee(\nu)-\mathbf W(\nu)\mathbf W^\vee(\mu)\Bigr)\,,
\]
and the Kirchhoff equations on $e(3)$ (\ref{Cl-eqm}) one gets the
following evolutionary equations
\bq\label{eq-xieta}\begin{array}{c}
           \dot{ \eta}= \left(\mathbf
C_{11}\eta+\mathbf C_{12}\xi\right)\times\eta+ \left(\mathbf
C_{21}\eta +\mathbf C_{22}\xi\right)\times\xi, \\
\\
           \dot{ \xi}= \left(\mathbf C_{11}\eta+\mathbf
C_{12}\xi\right)\times\xi+ \left(\mathbf C_{21}\eta+\mathbf
C_{22}\xi\right)\times\eta,
          \end{array}
\eq
where
\ben
\left(%
\begin{array}{cc}
 \mathbf C_{11} & \mathbf C_{12} \\
 \mathbf C_{21} & \mathbf C_{22}
\end{array}%
\right)&=&\left(%
\begin{array}{cc}
 \mathbf Z^{-1}\left[\mathbf W(\mu)+\mathbf W(\nu)\right] & 0 \\
 0 & \mathbf Z^{-1}\left[\mathbf W(\mu)-\mathbf W(\nu)\right] \\
\end{array}%
\right)
\left(%
\begin{array}{cc}
 -\mathbf W_+ & \mathbf W_- \\
 -\mathbf W_+ & \mathbf W_- \\
\end{array}%
\right),. \nn
\en
By definition we have $\mathbf C_{11}\mathbf C_{22}-\mathbf
C_{12}\mathbf C_{21}=0$.

Let us postulate the following brackets between $\xi$ and $\eta$
\begin{equation}\label{dso4}
\,\qquad
\bigl\{\eta_i\,,\eta_j\,\bigr\}=\varepsilon_{ijk}(\eta_k+\gamma\xi_k)\,,
\qquad
\bigl\{\eta_i\,,\xi_j\,\bigr\}=\varepsilon_{ijk}(\xi_k+\gamma\eta_k)
\,, \qquad
\bigl\{\xi_i\,,\xi_j\,\bigr\}=\varepsilon_{ijk}(\eta_k+\gamma\xi_k)\,,
\end{equation}
which coincide with the Lie-Poisson brackets on $so(4)$ (\ref{2o3})
after the following change of variables
\[
S_i=\dfrac{\eta_i-\xi_i}{2(1-\gamma)}\,,\qquad
T_i=\dfrac{\eta_i+\xi_i}{2(1+\gamma)}\,,
\]
here $\gamma$ is arbitrary.

The equations of motion (\ref{eq-xieta}) are Hamiltonian equations
with respect to the brackets (\ref{dso4}) and the following
quadratic Hamilton function
\[ {\mathscr H}=(\widetilde{\mathbf C}_{11} \eta, \eta)+(\widetilde{\mathbf C}_{12} \xi,
\eta)-(\widetilde{\mathbf C}_{22} \xi, \xi)\,,
\]
where
\[
\left(\begin{array}{cc}
 \mathbf C_{11} & \mathbf C_{12} \\
 \mathbf C_{21} & \mathbf C_{22}
\end{array}%
\right)=\left(\begin{array}{cc}
 1 & \gamma \\
 \gamma & 1
\end{array}%
\right)\left(\begin{array}{cc}
 \widetilde{\mathbf C}_{11} & \widetilde{\mathbf C}_{12} \\
 \widetilde{\mathbf C}_{12} & \widetilde{\mathbf C}_{22}
\end{array}%
\right).
\]
So, in fact K\"otter constructed two parametric transformation
$f_{\mu\nu}: so(4) \to e(3)$ defined by relations (\ref{gen-xieta})
that identifies the Clebsch system on $e(3)$ with the
Schottky--Manakov system on $so(4)$, it means that after mapping
$f_{\mu\nu}$ the corresponding
 Kirchhoff equations on $e(3)$ (\ref{Cl-eqm}) and on
 $so(4)$ (\ref{eq-xieta}) coincide.

If $a=b$ then $\gamma=0$ and the K\"otter mapping $f_{\mu\nu}$
(\ref{gen-xieta}) is equivalent to the twisted Poisson map $f_\mu$
(\ref{map-Cl}) (see Remark (\ref{kott-bob})). In generic case map
$f_{\mu\nu}$ (\ref{gen-xieta}) is the twisted Poisson map if
$a(\mu,\nu)$ and $b(\nu,\mu)$ are numerical functions. In order to
prove it we have to check that transformation $f_{\mu\nu}$
(\ref{gen-xieta}) gives rise to the second Poisson brackets
$\{.\,,.\}^*$ on $e(3)$ and $so(4)$, which are compatible with
initial ones.

According to \cite{shott91,kot92} (see also Remarks (\ref{rem-Neum})
and (\ref{rem-Neu})) equations (\ref{eq-xieta}) may be integrated in
elliptic coordinates if one of the Casimir elements on $so(4)$ is
equal to zero
\[( \xi, \eta)=0\,,\qquad \xi^2+\eta^2=const.\]
However in our case we have
\bq\label{inner-xieta} ( \xi, \eta)=a^2(\mu,\nu)\bfit
t^2(\mu)-b^2(\mu,\nu)\bfit t^2(\nu)\,, \qquad
\xi^2+\eta^2=2a^2(\mu,\nu)\bfit t^2(\mu)+2b^2(\mu,\nu)\bfit
t^2(\nu).
\eq
Using the two parametric K\"otter mapping $f_{\mu\nu}$
(\ref{gen-xieta}) the necessary relation $( \xi, \eta)=0$ may be
achieved if one puts
 $\mu=\mathfrak s_i$ and $\nu=\mathfrak s_k$, where $\mathfrak
 s_{i,k}$ are any of the roots of the equation
\[
\bfit t^2(\mathfrak s_j)\equiv\det {\mathscr L}_e(\mathfrak
s_j)=0\,,\qquad j=1,2,3,4.
\]
Of course, these substitutions destroy the Poissonity of the mapping
$f_{\mu\nu}$. As a sequence, substituting $\mu,\nu=\mathfrak
s_1,\mathfrak s_2$ and $\mu,\nu=\mathfrak s_3,\mathfrak s_4$
conversely into (\ref{gen-xieta}) we can construct two different
pair of orthogonal vectors
\[
\boldsymbol \xi=\left. \xi\right|_{\mu=\mathfrak s_1,\nu=\mathfrak
s_2}\,,\quad\boldsymbol \eta= \left.\eta\right|_{\mu=\mathfrak
s_1,\nu=\mathfrak s_2}\,,\qquad\mbox{\rm and}\qquad
\tilde{\boldsymbol \xi}=\left.\xi\right|_{\mu=\mathfrak
s_3,\nu=\mathfrak s_4}\,,\quad\tilde{\boldsymbol \eta}= \left.\eta
\right|_{\mu=\mathfrak s_3,\nu=\mathfrak s_4}\,
\]
that satisfy dynamical equations (\ref{eq-xieta}) and algebraic
relations
\bq\label{k-prop}
\boldsymbol \xi^2+\boldsymbol \eta^2=0\,,\quad(\boldsymbol
\xi,\boldsymbol \eta)=0,\qquad\mbox{\rm and}\qquad
\tilde{\boldsymbol \xi}^2+\tilde{\boldsymbol \eta}^2=0,\quad
(\tilde{\boldsymbol \xi},\tilde{\boldsymbol \eta})=0\,.
\eq
Using relations (\ref{inv-relK}) it is easy to prove that vectors
$\boldsymbol \xi,\boldsymbol \eta$ and $\tilde{\boldsymbol
\xi},\tilde{\boldsymbol \eta}$ are linearly dependent
\bq\label{lindep}
\left(\begin{array}{c}
 \tilde{\boldsymbol \xi} \\
 \tilde{\boldsymbol \eta}
\end{array}\right)=\mathbf U\left(%
\begin{array}{rr}
\widetilde{\mathbf W}_+ & -\mathbf W_+ \\
\widetilde{\mathbf W}_- & -\mathbf W_-
\end{array}%
\right)_{
 \mu=\mathfrak s_3\,
 \nu=\mathfrak s_4
}
\left(%
\begin{array}{rr}
\mathbf W_- & -\mathbf W_+\\
\widetilde{\mathbf W}_- & -\widetilde{\mathbf W}_+
\end{array}%
\right)_{\mu=\mathfrak s_1\,\nu=\mathfrak s_2}\left(\begin{array}{c}
 {\boldsymbol \xi} \\
 {\boldsymbol \eta}
\end{array}\right)
\,,
\eq
where
\[
\mathbf U=\mathbf Z(\mathfrak s_3,\mathfrak s_4)\mathbf
Z^{-1}(\mathfrak s_1,\mathfrak s_2)\Bigl(\mathbf
W_-\widetilde{\mathbf W}_+-\mathbf W_+\widetilde{\mathbf
W}_-\Bigr)^{-1}(\mathfrak s_3,\mathfrak s_4);
\]
Inserting (\ref{lindep}) into (\ref{k-prop}) we obtain four
algebraic equations, which relate six dynamical variables $\xi_i$
and $\eta_i$. These equations have non-trivial solutions
$\xi_i=g_i(\eta_1,\eta_2,\eta_3)$ for the special choice of the
functions $a(\mu,\nu)$ and $b(\mu,\nu)$ (\ref{gen-xieta}) only.

For instance, K\"otter uses the following functions
\[
a(\mu,\nu)=\frac{1}{\sqrt{\psi'(\mu)}},\qquad
b(\mu,\nu)=i\frac{1}{\sqrt{\psi'(\nu)}},
\]
 where
 \[\psi(\mu)=(\mu-\mathfrak s_1)(\mu-\mathfrak s_2)(\mu-\mathfrak s_3)(\mu-\mathfrak
 s_4)\,,\qquad
\psi'(\mu)=\frac{d \psi(\mu)}{d\mu}\,.
\]
In this case relation
\[
\mathbf W_\pm(\mathfrak s_1,\mathfrak s_2)\mathbf
W^{-1}_\pm(\mathfrak s_3,\mathfrak s_4)=\widetilde{\mathbf
W}_\pm(\mathfrak s_1,\mathfrak s_2)\widetilde{\mathbf
W}^{-1}_\pm(\mathfrak s_3,\mathfrak s_4)
\]
allows us to reduce equations (\ref{lindep}) to more simple form
$\tilde{\boldsymbol \xi}=\mathbf D\boldsymbol \xi$ and
$\tilde{\boldsymbol \eta}=\mathbf D^{-1}\boldsymbol \eta$, such that
four algebraic equations (\ref{k-prop}) give rise to three
independent equations only
\bq\label{sys-26} \sum_{j=1}^3(\xi_j^2+\eta_j^2)=0,
\qquad \sum_{j=1}^3 \xi_j\eta_j=0,\qquad \sum_{j=1}^3
\left(d_j^2\xi_j^2+\frac{\eta_j^2}{d_j^2}\right)=0\,.
\eq
Here $d_j$ are diagonal elements of the matrix $\mathbf D=\mathbf
W_+(\mathfrak s_3,\mathfrak s_4)\mathbf W^{-1}_+(\mathfrak
s_1,\mathfrak s_2)$
\[
d_j== \frac{\widetilde{w}_j(\mathfrak
s_4)+i\widetilde{w}_j(\mathfrak s_3)} {\widetilde{w}_j(\mathfrak
s_1)+i\widetilde{w}_j(\mathfrak s_2)},\qquad
\widetilde{w}_j(\mu)=\frac{{w_j(\mu)}}{\sqrt{\psi'(\mu)}}\,.
\]
Matrix $\mathbf D$ obeys the property $\mathbf D^*\mathbf D=-1$,
where $\mathbf D^*=\mathbf W_-(\mathfrak s_3,\mathfrak s_4)\mathbf
W^{-1}_-(\mathfrak s_1,\mathfrak s_2)$ is its Hermitian conjugate.

Introduce elliptic coordinates $\mathpzc z_{1,2}$ as the roots of
equation
\bq\label{cl-ell}
\mathpzc A\,\frac{(z-\mathpzc z_1)(z-\mathpzc z_2)}{\chi(z)}=
\frac{\eta_1^2}{z-d_1^2}
+\frac{\eta_2^2}{z-d_2^2}+\frac{\eta_3^2}{z-d_3^2}=0,
\eq
where $\chi(z)=(z-d_1^2)(z-d_2^2)(z-d_3^2)$ and $\mathpzc
A=\eta_1^2+\eta_2^2+\eta_3^2$. From equations (\ref{sys-26}) and
(\ref{cl-ell}) one gets
\[
\eta_j=\sqrt{\mathpzc A\,\frac{(\mathpzc z_1-d_j^2)(\mathpzc
z_2-d_j^2)}{\chi'(d_j^2)}},\qquad
\xi_j=\frac{\eta_j}{d_1d_2d_3}\frac{\sqrt{\mathpzc z_1\mathpzc
z_2}}{\mathpzc z_1-\mathpzc z_2}\left( \frac{\sqrt{P_5(\mathpzc
z_1)}}{\mathpzc z_1(\mathpzc z_1-d_j^2)} -\frac{\sqrt{P_5(\mathpzc
z_2)}}{\mathpzc z_2(\mathpzc z_2-d_j^2)}\right)\,,
\]
where $P_5(z)$ is a fifth order polynomial
\bq
\label{P_5(z)}
P_5(z)=z(z-d_1^2d_2^2d_3^2)\,\chi(z)=z(z-d_1^2d_2^2d_3^2)(z-d_1^2)(z-d_2^2)(z-d_3^2).
\eq
By definition (\ref{cl-ell}) velocities of the coordinates $\mathpzc
z_{1,2}$ are given by
\[
(-1)^j({\mathpzc z_1-\mathpzc z_2})\dot{\mathpzc
z}_j=\frac{2\chi(z_j)}{\mathpzc
A}\sum_{k=1}^3\frac{\eta_k\dot{\eta}_k}{\mathpzc z_j-d_k^2}\,,\qquad
j=1,2.
\]
Excluding $\dot \eta_j$ from equations of motion (\ref{eq-xieta})
one gets
\bq\label{sep_Kott}
(-1)^j({\mathpzc z_1-\mathpzc z_2})\dot{\mathpzc z}_j={(d\mathpzc
w_1\mathpzc z_j+d\mathpzc w_2)\sqrt{P_5(\mathpzc z_j)\,}}\,,\qquad
j=1,2,
\eq
where $d\mathpzc w_{1,2}$ are values of the constants of motion,
which we reproduce in the K\"otter form
\ben
d\mathpzc w_1&=&2\sum_{k=1}^3\frac{\eta_k\dot{\eta}_k}{\mathpzc
z_1-z_2}\left(
 \frac{\sqrt{P_5(\mathpzc z_1)}}{\mathpzc
z_1(\mathpzc z_1-d_k^2)(\mathpzc z_1-d_1^2d_2^2d_3^2)}
-\frac{\sqrt{P_5(\mathpzc z_2)}}{\mathpzc z_2(\mathpzc
z_2-d_k^2)(\mathpzc z_2-d_1^2d_2^2d_3^2)} \right)\nn\\
\label{dw}\\
d\mathpzc w_2&=&2\sum_{k=1}^3\frac{\eta_k\dot{\eta}_k}{\mathpzc
z_1-z_2}\left(
 \frac{\sqrt{P_5(\mathpzc z_1)}}{(\mathpzc z_1-d_k^2)(\mathpzc z_1-d_1^2d_2^2d_3^2)}
-\frac{\sqrt{P_5(\mathpzc z_2)}}{(\mathpzc z_2-d_k^2)(\mathpzc
z_2-d_1^2d_2^2d_3^2)} \right)\nn
\en
(see equation (56) in \cite{kot92}). Applying the Abel-Jacobi
inversion theorem to the equations (\ref{sep_Kott}) K\"otter then
expressed initial variables $\bfit l$ and $\bfit p$ in
theta-functions.

On the next step we have to rewrite coefficients of  $P_5(z)$
(\ref{P_5(z)}) and integrals $d\mathpzc w_{1,2}$ (\ref{dw}) as
functions of the initial Clebsch integrals  and have to prove that
the K\"otter variables $\mathpzc z_{1,2}$ (\ref{cl-ell}) are in
involution with respect to initial Poisson brackets for the
Clebsch system. It is not easy task since $\{\eta_i,\eta_j\}\neq
0$ and $\{d_k,\eta_j\}\neq 0$. We suppose that or $\mathpzc
z_{1,2}$ commute or their commutativity may be restored by using
another functions $a(\mu,\nu)$ and $b(\mu,\nu)$ in
(\ref{gen-xieta}) because the K\"otter solution in theta-functions
was reproduced in framework of the finite-band integration
technique \cite{bbe94}. Recall that in \cite{kow89} Kowalevski
used non-commutative variables too (see Remark (\ref{Kow-rem}).

On the other hand, the matrix ${\mathscr L}_e(\mu)$ (\ref{Lax-K})
belongs to the family of the Lax matrices for  elliptic or $XYZ$
Gaudin magnet. The separated variables for the generic Gaudin
model were constructed in \cite{skl98} in classical and quantum
mechanics. So, we can construct the separated variables for the
Clebsch system and, therefore, for the Kowalevski gyrostat using
the Sklyanin method, which is free from the difficulties of the
K\"otter approach.

\section{Conclusion}

Our treatment shows that including  gyrostatic term into the
Hamiltonian of the Kowalevski top we arrive at the essentially
more complicate dynamic system. Recall once more that quantum
corrections to the Kowalevski top looks similar to the gyrostatic
term.

The main result of this paper is that we establish one to one
correspondence between the Kowalevski gyrostat and the Clebsch
system. It allows us to construct solution of the gyrostat problem
using various known solutions of the Clebsch model. The
separation of variables for the Clebsch system that is unknown now
becomes a matter of a primary importance.

The similar correspondence may be obtained for the Kowalevski
gyrostat on $so(4)$ algebra, which is equivalent to the generalized
Kowalevski gyrostat on $e(3)$ \cite{kst03}. For the $so(4)$ top
initial vector $\bfit l_{so(4)}$ was obtained in \cite{komkuz}. In
the gyrostat case we have to substitute it by the rule (\ref{lg}).

Authors thank V V Sokolov for revival our interest to the Kowalevski
gyrostat and for useful discussions. The research of AVT was
partially supported by RFBR grant 02-01-00888.

\end{document}